\documentclass[aps,showpacs,amsmath,preprint,amssymb,12pt]{revtex4}
\usepackage{graphicx}

\def\lam{{\lambda}}
\def\gam{{\gamma}}

\def\tilU{{\tilde{U}}}
\def\tU{{\tilde{U}}}

\begin{document}


\title{Boosted Black Holes on Kaluza-Klein Bubbles }

\author{Hideo Iguchi${}^{1}$, Takashi Mishima${}^{1}$ and Shinya Tomizawa${}^{2}$} 

\affiliation{
${}^{1}$ Laboratory of Physics,~College of Science and Technology,~
Nihon University,\\ Narashinodai,~Funabashi,~Chiba 274-8501,~Japan \\
${}^{2}$ Department of Mathematics and Physics, Graduate School of Science, Osaka City University, 3-3-138 Sugimoto, Sumiyoshi,
Osaka~558-8585,~Japan
}

\date{\today}

\begin{abstract}
We construct an exact stationary solution of black hole -- bubble sequence in the five dimensional Kaluza-Klein theory by using solitonic solution generating techniques. The solution describes two boosted black holes with topology $S^3$ on a Kaluza-Klein bubble and has a linear momentum component in the compactified direction. The ADM mass and the linear momentum depend on the two boosted velocity parameters of black holes. In the effective four dimensional theory, the solution has an electric charge which is proportional to the linear momentum.  The solution includes the static solution found by Elvang and Horowitz and a limit of single boosted black string.
\end{abstract}

\preprint{OCU-PHYS 268}
\preprint{AP-GR 44}
\maketitle

\section{Introduction}

Kaluza-Klein (KK) theory is the five dimensional theory of gravity which unifies Einstein's four-dimensional theory of gravity and Maxwell's electromagnetic theory
\cite{{KK},{Overduin:1998pn}}.
The spacetime is asymptotically the product of the four-dimensional Minkowski spacetime ${\cal M}^{3,1}$ and a circle $S^1$.
The extra dimension with $S^1$ is compactified too small for us to observe it.
This type of compactification of the extra dimensions are also extended to the supergravity 
theories and the superstrings. 

The studies on black holes in the KK theory have attracted much attention since they admit much richer structures than asymptotically flat higher dimensional black holes. For example, such black holes can have the horizons with the topologies of the squashed $S^3$ and the lens space $L(2;1)=S^3/{\mathbb Z}_2$~\cite{IM,Ishihara}.  
Another exciting aspects of the KK theory are the existence of KK bubbles. In Ref.~\cite{Elvang:2004iz}, a large class of five- and six-dimensional static solutions which describe the sequences of the black holes and KK bubbles were constructed and analyzed. 
The KK bubble was first found by Witten as the end state of the KK vacuum decay \cite{Witten}. 
The first solution of the sequence of black holes and bubbles is
the combination of the static black hole and KK bubble \cite{weyl}.
Elvang and Horowitz found and analyzed the two black holes sitting on the KK bubble
\cite{EH}. The static equilibrium of the spacetime is maintained by the existence of the KK bubbles which balance the attractive force of black holes.

In the previous article, we obtained the new five-dimensional vacuum solution of rotating black holes on the KK bubble \cite{Tomizawa:2007mz}. This solution is the extension of the static solution found by Elvang and Horowitz to a stationary solution, which has an ADM mass and an ADM angular momentum. We used two different types of solution generating methods to obtain the solution. 
One is called B\"acklund transformation \cite{Harrison,Neugebauer},
which is basically the technique to generate 
a new solution of the Ernst equation. The other is the inverse scattering technique,
which Belinski and Zakharov~\cite{Belinskii} developed
as an another type of solution-generating technique. 
In this several years, these techniques have been applied to generate and to reproduce
five-dimensional black hole solutions with asymptotically flatness
\cite{Mishima,MI2,MI3,Koikawa,Tomizawa,Azuma,Pomeransky:2005sj,Tomizawa2,Pomeransky2,EF}. The relation between these two methods was examined
in the context of the five-dimensional spacetime \cite{Tomizawa3}. It was shown that the two-solitonic solutions generated from an arbitrary diagonal seed by the B\"acklund transformation coincide with those with a single angular momentum generated from the same seed by the inverse scattering method. 

In this article we generate another type of stationary solution which describes boosted black holes on the KK-bubble as a vacuum solution in the five-dimensional Einstein equations by using both solitonic methods. It should be noted that this solution cannot be generated by the simple boost transformation of the static solution because the boosted solution always has closed timelike curves around the bubble. The solution has a linear momentum in the compact direction and does not have an ADM angular momentum.
This is the reason why we call the solution boosted black holes on KK bubble. In the four dimensional effective theory, this solution has an electric charge which is proportional to the linear momentum. There is a limit of single black hole without KK bubble in this solution. This limiting solution exactly corresponds to the simply boosted black string \cite{Dobiasch:1981vh,Gibbons:1985ac,Cvetic:1995sz,Chamblin:1996kw} whose thermodynamical properties are studied recently \cite{Kastor:2007wr}. The solution with two momentum components will be generated by the inverse scattering method

This article is organized as follows: In Sec.\ref{sec:solution}, we give a new solution generated by the solitonic methods. We introduce only the construction by the B\"aclund transformation in this section, while the other construction is briefly mentioned 
in Appendix \ref{app:ISM}. In Sec.\ref{sec:properties}, we investigate the properties of the solution. In Sec.\ref{sec:final}, we give the summary and discussion of this article. In Appendix, we give the solution generated by the inverse scattering method and the relation between these solutions.

\section{Solutions}
\label{sec:solution}
At first we briefly 
present the solution obtained by the B\"acklund transformation
which was applied the five dimensional case \cite{Mishima}.
Using this method, we can generate axially symmetric solutions 
of five-dimensional vacuum Einstein equations. 
See \cite{Mishima} for the detail of the solution generating method.

We start from the following form of a seed static metric
\begin{equation}
ds^2 =e^{-T^{(0)}}\left[
       -e^{S^{(0)}}(dt)^2
       +e^{-S^{(0)}}\rho^2(d\phi)^2 
   +e^{2\gamma^{(0)}-S^{(0)}}\left(d\rho^2+dz^2\right) \right]
  +e^{2T^{(0)}}(d\psi)^2,
\end{equation}
with seed functions
\begin{eqnarray}
S^{(0)}&=& {U}_{\lambda \sigma} - \tilde{U}_{\eta_1\sigma}
           +2\tilde{U}_{\eta_2\sigma}\nonumber \\
       &=&  -\tilde{U}_{\lambda \sigma} - \tilde{U}_{\eta_1\sigma}
           +2\tilde{U}_{\eta_2\sigma}+\ln \rho      \label{eq:seed_2}\\
T^{(0)} &=& {U}_{\lambda \sigma} + \tilde{U}_{\eta_1\sigma}\nonumber \\
        &=& -\tilde{U}_{\lambda \sigma} + \tilde{U}_{\eta_1\sigma}+\ln \rho,
\end{eqnarray}
where we assume $\eta_1<\eta_2<\lambda<1$. 
The function $U_{d}$ is defined as $U_{d}:=\frac{1}{2}\ln\left[\,R_{d}-(z-d)\,\right]$
and the function $\tilU_{d}$ is defined as $\tilU_{d}:=\frac{1}{2}\ln\left[\,R_{d}+(z-d)\,\right]$ where $R_d:=\sqrt{\rho^2+(z-d)^2}$.
Here we take the coordinate $\phi$ as a Kaluza-Klein compactified direction. As explained later, the solitonic solution has two event horizons at $\eta_1\sigma \le z \le \eta_2\sigma$ and $-\sigma \le z \le \lambda\sigma$ and a Kaluza-Klein bubble at  
$\eta_2\sigma \le z \le -\sigma$, where the Kaluza-Klein circles shrink to zero. The metric of the solitonic solution can be written in the 
following form
\begin{eqnarray}
ds^2 &=&e^{-T}\left[
       -e^{S}(dt-\omega d\phi)^2
       +e^{-S}\rho^2(d\phi)^2 
+e^{2\gamma-S}\left(d\rho^2+dz^2\right) \right]
  +e^{2T}(d\psi)^2.
  \label{MBmetric}
\end{eqnarray}
The function $T$ is derived from
the seed functions
\begin{equation}
T=-\tilde{U}_{\lambda \sigma} + \tilde{U}_{\eta_1\sigma}+\ln \rho.
\end{equation}
The other metric functions for the five-dimensional metric 
 (\ref{MBmetric}) are obtained
by using the formulas shown by \cite{Castejon-Amenedo:1990b}, 
\begin{eqnarray}
e^{S}&=&e^{S^{(0)}}\frac{A}{B} ,  \label{e^S} \\
\omega&=&2\sigma e^{-S^{(0)}}\frac{C}{A}-C_1 \label{omega} ,    \\
e^{2\gamma}&=&C_2(x^2-1)^{-1}A
                e^{2\gamma'}, \label{e_gamma}
\end{eqnarray}
where $C_1$ and $C_2$ are constants and
$A$, $B$ and $C$ are given by
\begin{eqnarray*}
 A &=& \frac{1}{(2\sigma)^2}
  \left\{\left(e^{2\tU_{-\sigma}}+e^{2U_{\sigma}}\right)
    \left(e^{2\tU_{\sigma}}+e^{2U_{-\sigma}}\right) (1+ab)^2 
        -\left(e^{2\tU_{-\sigma}}-e^{2\tU_{\sigma}}\right)
    \left(e^{2U_{\sigma}}-e^{2U_{-\sigma}}\right) (b-a)^2\right\}, \\
 B &=& \frac{1}{(2\sigma)^2} 
  \left\{\left[\left(e^{2\tU_{-\sigma}}+e^{2U_{\sigma}}\right)
    +\left(e^{2\tU_{\sigma}}+e^{2U_{-\sigma}}\right)ab \right]^2
    +\left[\left(e^{2\tU_{-\sigma}}-e^{2\tU_{\sigma}}\right)a
    -\left(e^{2U_{\sigma}}-e^{2U_{-\sigma}}\right)b\right]^2\right\},
 \\
 C &=& \frac{1}{(2\sigma)^3} 
     \left\{
         \left(e^{2\tU_{-\sigma}}+e^{2U_{\sigma}}\right)
         \left(e^{2\tU_{\sigma}}+e^{2U_{-\sigma}}\right) (1+ab)
       \left(  \left(e^{2U_{\sigma}}-e^{2U_{-\sigma}}\right) b
      -\left(e^{2\tU_{-\sigma}}-e^{2\tU_{\sigma}}\right) a \right)
     \right. \nonumber \\  && \left. \hspace{0.cm}
     +\left(e^{2\tU_{-\sigma}}-e^{2\tU_{\sigma}}\right)
       \left(e^{2U_{\sigma}}-e^{2U_{-\sigma}}\right) (b-a)
      \left(\left(e^{2\tU_{-\sigma}}+e^{2U_{\sigma}}\right)
        - \left(e^{2\tU_{\sigma}}+e^{2U_{-\sigma}}\right) ab\right)
     \right\}.
\end{eqnarray*}
The functions $a$ and $b$, which are auxiliary potential to obtain 
the new Ernst potential by the transformation,
are given by
\begin{eqnarray}
a
&=& \alpha
   \sqrt{\frac{(e^{2\tU_{-\sigma}}+e^{2U_{\sigma}})(e^{2U_{\sigma}}-e^{2U_{-\sigma}})}{(e^{2\tU_{\sigma}}+e^{2U_{-\sigma}})(e^{2\tU_{-\sigma}}-e^{2\tU_{\sigma}})}} 
   \frac{e^{\tilU_{\lam\sigma}}}{e^{2U_{\sigma}}+e^{2\tilU_{\lam\sigma}}}
   \frac{e^{\tilU_{\eta_1\sigma}}}
        {e^{2U_{\sigma}}+e^{2\tilU_{\eta_1\sigma}}}
   \left(\frac{e^{2U_{\sigma}}+e^{2\tilU_{\eta_2\sigma}}}
        {e^{\tilU_{\eta_2\sigma}}}\right)^2, \\
b
&=& \beta
   \sqrt{\frac{(e^{2\tU_{-\sigma}}+e^{2U_{\sigma}})(e^{2\tU_{-\sigma}}-e^{2\tU_{\sigma}})}{(e^{2\tU_{\sigma}}+e^{2U_{-\sigma}})(e^{2U_{\sigma}}-e^{2U_{-\sigma}})}}
   \frac{e^{2U_{-\sigma}}+e^{2\tilU_{\lam\sigma}}}{e^{\tilU_{\lam\sigma}}}
   \frac{e^{2U_{-\sigma}}+e^{2\tilU_{\eta_1\sigma}}}
        {e^{\tilU_{\eta_1\sigma}}}
   \left(\frac{e^{\tilU_{\eta_2\sigma}}}
        {e^{2U_{-\sigma}}+e^{2\tilU_{\eta_2\sigma}}}\right)^2.
\end{eqnarray}
In addition the function $\gamma'$ is obtained as
\begin{eqnarray}
\gam' &=& \gam'_{\sigma,\sigma}+\gam'_{-\sigma,-\sigma}
+\gam'_{\lambda\sigma,\lambda\sigma}+\gam'_{\eta_1\sigma,\eta_1\sigma}
+\gam'_{\eta_2\sigma,\eta_2\sigma}
 \nonumber \\
&&-2\gam'_{\sigma,-\sigma}
-\gam'_{\sigma,\lambda\sigma}
-\gam'_{\sigma,\eta_1\sigma}+2\gam'_{\sigma,\eta_2\sigma}
+\gam'_{-\sigma,\lambda\sigma}
+\gam'_{-\sigma,\eta_1\sigma}-2\gam'_{-\sigma,\eta_2\sigma} \nonumber \\
&&-\gam'_{\lambda\sigma,\eta_1\sigma} -\gam'_{\lambda\sigma,\eta_2\sigma} 
-\gam'_{\eta_1\sigma,\eta_2\sigma} \nonumber \\
&& +\tilde{U}_{\sigma}-\tilde{U}_{-\sigma}-2\tilde{U}_{\lambda\sigma}
   +\tilde{U}_{\eta_1\sigma}+\tilde{U}_{\eta_2\sigma} +\ln\rho,
\end{eqnarray}
where
\begin{equation}
\gam'_{cd}=\frac{1}{2}\tU_{c}+\frac{1}{2}\tU_{d}
           -\frac{1}{4}\ln [R_cR_d+(z-c)(z-d)+\rho^2]. \label{gam'}
\end{equation}
The constants $C_1$ and $C_2$ are chosen as follows
\begin{eqnarray}
C_1=0,\quad C_2=\frac{1}{(1+\alpha\beta)^2},\label{eq:asymp-con}
\end{eqnarray}
to avoid the global boost of the spacetime and to set the period of
$\psi$ to $2\pi$, respectively.
Also the integration constants $\alpha$ and $\beta$ should be decided as 
\begin{eqnarray}
\alpha^2 = \frac{(1-\lambda)(1-\eta_1)}{(1-\eta_2)^2},  \quad
\beta = 0,
\label{eq:alpha} 
\end{eqnarray} 
to remove the singularity at $z=\sigma$ on $z$-axis and closed timelike
curves around the bubble, respectively.


\section{Properties}\label{sec:properties}

Next, we investigate the properties of the solution satisfying the conditions (\ref{eq:asymp-con}) and (\ref{eq:alpha}). In particular, we study the asymptotic structure, the geometry of two black hole horizons and a bubble and the limits of the static case and the single boosted black hole.


\subsection{Asymptotic structure}
In order to investigate the asymptotic structure of the solution, let us introduce the coordinate $(r,\theta)$ defined as
\begin{eqnarray}
\rho=r\sin\theta,\quad z=r\cos \theta,
\end{eqnarray}
where $0\le\theta < 2\pi$ and $r$ is a four-dimensional radial coordinate in the neighborhood of the spatial infinity.
For the large $r\to\infty$, each component behaves as
\begin{equation}
g_{tt} \simeq -1 + \frac{(2-\eta_1 + \eta_2)\sigma}{r},\label{eq:a1}
\end{equation}
\begin{equation}
g_{\rho\rho}=g_{zz} \simeq 1 + \frac{(\lambda-\eta_1)\sigma}{r},\label{eq:a2}
\end{equation}
\begin{equation}
g_{t\phi} \simeq -\frac{2\alpha\sigma}{r},   \label{eq:a3}
\end{equation}
\begin{equation}
g_{\phi\phi} \simeq 1 + \frac{(2+\eta_2 -\lambda)\sigma}{r},\label{eq:a4}
\end{equation}
\begin{equation}
g_{\psi\psi} \simeq r^2 \sin^2 \theta \left(1-\frac{(\eta_1-\lambda)\sigma}{r}\right).\label{eq:a5}
\end{equation}
Hence, the leading order of the metric takes the form
\begin{eqnarray}
ds^2\simeq -dt^2+dr^2+r^2(d\theta^2+\sin^2\theta d\psi^2)+d\phi^2.
\end{eqnarray}
Therefore, the spacetime has the asymptotic structure of the direct product of the four-dimensional Minkowski spacetime and $S^1$. The $S^1$ at infinity is parameterized by $\phi$ and the size $\Delta\phi$ is given in \ref{sec:rod}.

\subsection{Mass and momentum}
Next, we compute the total mass and the linear momentum of the spacetime. It should be noted that since the asymptotic structure is ${\cal M}^{3,1}\times S^1$, the ADM mass and momentum are given by the surface integral over the spatial infinity with the topology of $S^2\times S^1$. In order to compute these quantities, we introduce asymptotic Cartesian coordinates $(x,y,z,\phi)$, where $x=\rho\cos\psi$ and $y=\rho\sin\psi$. Then, the ADM mass and momentum in the $\phi$ direction are given by
\begin{eqnarray}
{\cal M}_{\rm ADM}=\frac{1}{16\pi}\int_{S^2\times S^1}(\partial_j h_{ij}-\partial_i h_{jj})dS_i,
\label{eq:adm_mass_def}
\end{eqnarray}
\begin{eqnarray}
 {\cal P} = \frac{1}{16\pi}\int_{S^2\times S^1} \partial_i h_{0\phi}dS_i
\end{eqnarray}
respectively. Here $h_{\mu\nu}$ is deviation from the  five-dimensional flat metric
$\eta_{\mu\nu}$ near infinity,
\begin{eqnarray}
 g_{\mu\nu}=\eta_{\mu\nu}+h_{\mu\nu}.
\end{eqnarray}
The Latin index $i,j$ runs $x,y,z$ and $\phi$ and the Greek indeces $\mu,\nu,\alpha$ and $\beta$ label $t,x,y,z$ and $\phi$.
Then, the ADM mass of the solution is computed as
\begin{eqnarray}
{\cal M}_{ADM}=\frac{(\lambda-2\eta_1+\eta_2+2)\sigma}{4}\Delta\phi.
\label{eq:adm_mass}
\end{eqnarray}
It should be noted that the ADM mass is non-negative.
The linear momentum becomes
\begin{eqnarray}
{\cal P}=-\frac{\alpha\sigma}{2}\Delta\phi.
\end{eqnarray}
The electric charge of four dimensional effective theory is  proportional to the linear momentum per unit $\phi$ length. We define it as
\begin{equation}
 Q_4 = \alpha \sigma.
\label{eq:charge_def}
\end{equation}



\subsection{Black holes and bubble}\label{sec:rod}
\begin{figure}[htbp]
\begin{center}
\includegraphics[width=0.6\linewidth]{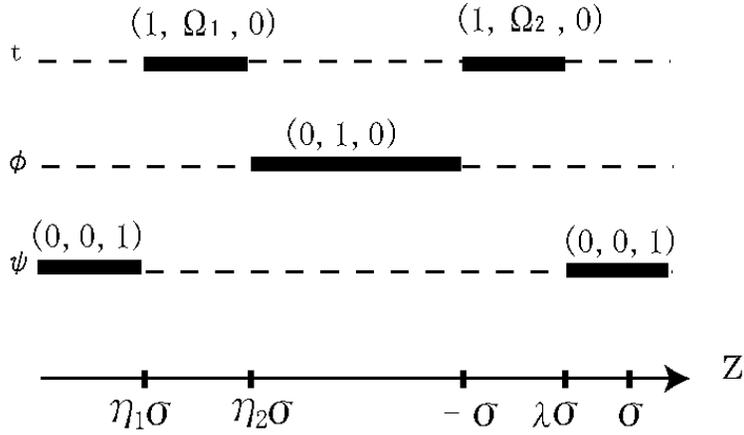}
\end{center}
\caption{Rod structure of boosted black holes on a Kaluza-Klein bubble. The finite timelike rods $[\eta_1\sigma,\eta_2\sigma]$ and $[-\sigma,\lambda\sigma]$ correspond to black holes with boost velocities $\Omega_1$ and $\Omega_2$, respectively. The finite spacelike rod $[\eta_2\sigma,-\sigma]$ denotes a Kaluza-Klein bubble, where Kaluza-Klein circles shrink to zero.}\label{fig:rod}
\end{figure}

Here, for the solution, we consider the rod structure developed by Harmark~\cite{Harmark} and Emparan and Reall~\cite{weyl}. 
The rod structure at $\rho=0$ is illustrated in FIG.\ref{fig:rod}.
(i) The finite timelike rod $[\eta_1\sigma,\eta_2\sigma]$ and $[-\sigma,\lambda\sigma]$ denote the locations of black hole horizons. These timelike rods have directions $v_1=(1,\Omega_1,0)$ and $v_2=(1,\Omega_2,0)$. We call $\Omega_1$ and $\Omega_2$ boost velocity parameters. These are given by
\begin{equation}
\Omega_1 = \frac{2\alpha}
             {1-\eta_1},
\end{equation} 
for $\eta_1\sigma < z < \eta_2 \sigma$ and
\begin{equation}
 \Omega_2 = \frac{\alpha(1-\eta_2)^2}
                 {2(1-\eta_1)},
\end{equation} 
for $-\sigma < z < \lambda\sigma.$ Here, it should be noted that $\Omega_1$ and $\Omega_2$ have the same signature. Therefore, two black holes are boosted along the same direction. 
(ii) The finite spacelike rod $[\eta_2\sigma,-\sigma]$ 
which corresponds to a Kaluza-Klein bubble has the direction $v=(0,1,0)$. In order to avoid conical singularity for $z\in[\eta_2\sigma,-\sigma]$ and $\rho=0$, $\phi$ has the periodicity of
\begin{eqnarray}
\frac{\Delta\phi}{2\pi} &=& \lim_{\rho\rightarrow \infty} \sqrt{\frac{\rho^2g_{\rho\rho}}{g_{\phi\phi}}} \nonumber\\
            &=& 2\sigma\frac{\eta_2+1}
                           {\eta_2-1}
             \sqrt{\frac{(\lambda-\eta_1)(\lambda-\eta_2)(\eta_1-1)}{\eta_1+1}}.
\end{eqnarray}
(iii) The semi-infinite spacelike rods $[-\infty,\eta_1\sigma]$ and $[\lambda\sigma,\infty]$ have the direction $v=(0,0,1)$. In order to avoid conical singularity,
$\psi$ has the periodicity of
\begin{equation}
\Delta\psi= 2\pi.
\end{equation}

Here, we write the induced metrics of the event horizons and the bubble. 
For $\eta_1\sigma<z<\eta_2\sigma$, the induced metric 
becomes
\begin{eqnarray}
 g_{tt}
 &=& \frac{\Omega_1^2(z-\sigma)(z+\sigma)(z-\eta_2\sigma)}{(z-\lambda\sigma)(z+\sigma)^2-\Omega_1^2(z-\eta_1\sigma)(z-\eta_2\sigma)^2},
\end{eqnarray}
\begin{eqnarray}
 g_{t\phi}
  &=&-\frac{\Omega_1(z-\sigma)(z+\sigma)(z-\eta_2\sigma)}{(z-\lambda\sigma)(z+\sigma)^2-\Omega_1^2(z-\eta_1\sigma)(z-\eta_2\sigma)^2},
\end{eqnarray}
\begin{eqnarray}
 g_{zz}
 &=& \frac{\sigma^2(1-\eta_1)(\eta_2-\eta_1)(\lambda-\eta_1)((z-\lambda\sigma)(z+\sigma)^2-\Omega_1^2(z-\eta_1\sigma)(z-\eta_2\sigma)^2)}
{(1+\eta_1)(z-\sigma)(z+\sigma)(z-\eta_1\sigma)(z-\eta_2\sigma)(z-\lambda\sigma)},
\end{eqnarray}
\begin{eqnarray}
 g_{\phi\phi}
&=& \frac{(z-\sigma)(z+\sigma)(z-\eta_2\sigma)}{(z-\lambda\sigma)(z+\sigma)^2-\Omega_1^2(z-\eta_1\sigma)(z-\eta_2\sigma)^2},
\end{eqnarray}
\begin{equation}
g_{\psi\psi}=-4(z-\eta_1\sigma)(z-\lambda\sigma),
\end{equation}
Since the $\phi$ circles shrink to zero at $z=\eta_2\sigma$ and $\psi$ circles shrink to zero at $z=\eta_1\sigma$, the spatial cross section of this black hole horizon is topologically $S^3$.
The area of the event horizon is 
\begin{eqnarray}
 A_1 &=& 8\pi^2 \sigma^2 \sqrt{\frac{(1-\eta_1)(\lambda-\eta_1)(\eta_2-\eta_1)^3}{-1-\eta_1}}
 \Delta \phi \nonumber \\
  &=& 16 \pi^2 
\sigma^3 \frac{(1-\eta_1)(-1-\eta_2)(\eta_2-\eta_1)^{\frac{3}{2}}(\lambda-\eta_1)
(\lambda-\eta_2)^{\frac{1}{2}}}{(-1-\eta_1)(1-\eta_2)}.
\end{eqnarray}

For $-\sigma<z<\lambda\sigma$, the induced metric takes the following form
\begin{eqnarray}
 g_{tt}
&=& \frac{\Omega_2^2(z-\sigma)(z+\sigma)(z-\eta_2\sigma)}
{(z-\lambda\sigma)(z-\eta_2\sigma)^2-\Omega_2^2(z+\sigma)^2(z-\eta_1\sigma)},
\end{eqnarray}
\begin{eqnarray}
 g_{t\phi}
 &=& -\frac{\Omega_2(z-\sigma)(z+\sigma)(z-\eta_2\sigma)}
{(z-\lambda\sigma)(z-\eta_2\sigma)^2-\Omega_2^2(z+\sigma)^2(z-\eta_1\sigma)},
\end{eqnarray}
\begin{eqnarray}
 g_{zz}
 &=& -\frac{4\sigma^2(1-\eta_1)(\eta_1-\lambda)(\eta_2-\lambda)((z-\lambda\sigma)(z-\eta_2\sigma)^2-\Omega_2^2(z+\sigma)^2(z-\eta_1\sigma))}
{(1-\eta_2)^2(1+\lambda)(z-\sigma)(z+\sigma)(z-\eta_1\sigma)(z-\eta_2\sigma)(z-\lambda\sigma)},
\end{eqnarray}
\begin{eqnarray}
 g_{\phi\phi}
&=&\frac{(z-\sigma)(z+\sigma)(z-\eta_2\sigma)}
{(z-\lambda\sigma)(z-\eta_2\sigma)^2-\Omega_2^2(z+\sigma)^2(z-\eta_1\sigma)},
\end{eqnarray}
\begin{equation}
g_{\psi\psi}=-4(z-\eta_1\sigma)(z-\lambda\sigma),
\end{equation}
Since the $\phi$ circles shrink to zero at $z=-\sigma$ and $\psi$ circles shrink to zero at $z=\lambda\sigma$, the spatial cross section of this black hole horizon is also topologically $S^3$.
The area of this event horizon is 
\begin{eqnarray}
 A_2 &=& 16\pi^2\sigma^2 \sqrt{(1-\eta_1)(\lambda-\eta_1)(\lambda-\eta_2)(\lambda+1)}
         \Delta \phi \nonumber \\
&=& 32 \pi^2 \sigma^3 \frac{(1-\eta_1)(-1-\eta_2)(\lambda-\eta_1)(\lambda-\eta_2)(\lambda+1)^{\frac{1}{2}}}
{(1-\eta_2)(-1-\eta_1)^{\frac{1}{2}}}.
\end{eqnarray}

For $\eta_2\sigma<z<-\sigma$, the induced metric on the bubble can be written in the form 
\begin{equation}
 g_{tt}=-\frac{(z+\sigma)(z-\eta_2\sigma)}{(z-\sigma)(z-\eta_1\sigma)},
\end{equation}
\begin{equation}
 g_{t\phi}
 =-\frac{\sigma^2\Omega_2\rho^2}{(z+\sigma)(z-\eta_2\sigma)(z-\lambda\sigma)},
\end{equation}
\begin{equation}
 g_{zz} =\sigma^2 \frac{(1-\eta_1)(1+\eta_2)^2(\lambda-\eta_1)(\lambda-\eta_2)(z-\sigma)}
{(1+\eta_1)(1-\eta_2)^2(z+\sigma)(z-\eta_2\sigma)(z-\lambda\sigma)},
\end{equation}
\begin{equation}
 g_{\phi\phi} = - \frac{(z-\sigma)\rho^2}{4(z-\eta_2\sigma)(z-\lambda\sigma)(z+\sigma)},
\end{equation}
\begin{equation}
g_{\psi\psi}=-4(z-\eta_1\sigma)(z-\lambda\sigma),
\end{equation}
The $\phi$ circle vanishes for $z\in [\eta_2\sigma,-\sigma]$ and $\rho=0$, which means that there exists a Kaluza-Klein bubble in this region. Since the $\psi$ circle does not vanish at $z=\eta_2\sigma$ and $z=\lambda\sigma$, this bubble on the time slice is topologically a cylinder $S^1\times R$. Therefore, there exist a Kaluza-Klein bubble between two boosted black holes with topology of $S^3$. 
The proper distance between the two black hols is
\begin{equation}
 s= \sigma \frac{(\eta_2+1)}{(\eta_2-1)}
    \sqrt{\frac{(\eta_1-1)(\lambda-\eta_1)(\lambda-\eta_2)}{(\eta_1+1)}}
    \int^{-\sigma}_{\eta_2\sigma}dz\sqrt{-\frac{z-\sigma}{(z+\sigma)(z-\eta_2\sigma)(z-\lambda\sigma)}}.
\end{equation}
The Kaluza-Klein bubble is significant to keep the balance of two black holes and
achieve the solution without any strut structures and singularities.
This property resembles that of the solution given by Elvang and Horowitz~\cite{EH}
and the extension of it with rotation \cite{Tomizawa:2007mz}. In the next subsection, we will show that the static limit of the solution coincides with the solution given by Elvang and Horowitz.

At the end of this subsection we rewrite the ADM mass (\ref{eq:adm_mass}) and the electric charge (\ref{eq:charge_def})
by using boost velocities $\Omega_1$ and $\Omega_2$.
The ADM mass is
\begin{eqnarray}
 \frac{M_{ADM}}{\Delta\phi}&=&\frac{(\lambda-\eta_1)\sigma}{2}
                            \left(1+\frac{1-\lambda}{2(\lambda-\eta_1)}\right)
                            +\frac{(1+\eta_2)\sigma}{4} \nonumber \\
 &=& m\left(1+\frac{\Omega_1\Omega_2}{2(1-\Omega_1\Omega_2)}\right)
     +m_0 \left(1-\sqrt{\frac{\Omega_1}{\Omega_2}}\right), \label{eq:adm_mass2}
\end{eqnarray}
where $m=\frac{(\lambda-\eta_1)\sigma}{2}$ and $m_0=\frac{\sigma}{2}$.
The eclectic charge is
\begin{equation}
 Q_4=\alpha\sigma = m \frac{\Omega_1}{1-\Omega_1\Omega_2}. \label{eq:charge}
\end{equation}

\subsection{Static case}
In this subsection, we consider the static case, which can be obtained by the choice of the parameter $\lambda= 1$. Then, from Eq. (\ref{eq:alpha}) we see that $\alpha$ vanishes. Let us define the parameters $\tilde a,\tilde b$ and $\tilde c$ as
\begin{eqnarray}
\tilde a=\frac{2\lambda+1-\eta_2}{2}\sigma,\quad 
\tilde b=\frac{-1-\eta_2}{2}\sigma,\quad 
\tilde c=\frac{-1+\eta_2-2\eta_1}{2}\sigma.
\end{eqnarray}
It should be noted that $\lambda=1$ is equal to the condition $\sigma=(\tilde a-\tilde b)/2$.
Furthermore, let us shift an origin of the $z$-coordinate such that $z\to\tilde z:=z-(\eta_2-\lambda)\sigma/2$.
Then, we obtain the metric
\begin{eqnarray}
ds^2&=&-\frac{(R_{\tilde b}-(\tilde z-\tilde b))(R_{-\tilde c}-(\tilde z+\tilde c))}{(R_{\tilde a}-(\tilde z-\tilde a))(R_{-\tilde b}-(\tilde z+\tilde b))}dt^2+(R_{\tilde a}-(\tilde z-\tilde a))(R_{-\tilde c}-(\tilde z+\tilde c))d\psi^2\nonumber\\
   & &+\frac{R_{-\tilde b}-(\tilde z+\tilde b)}{R_{\tilde b}-(\tilde z-\tilde b)}d\phi^2+\frac{Y_{\tilde a,-\tilde c}Y_{\tilde b,-\tilde b}}{4R_{\tilde a}R_{\tilde b}R_{-\tilde b}R_{-\tilde c}}\sqrt{\frac{Y_{\tilde a,\tilde b}Y_{-\tilde b,-\tilde c}}{Y_{\tilde a,-\tilde b}Y_{\tilde b,-\tilde c}}}\frac{R_{\tilde a}-({\tilde z}-\tilde a)}{R_{-\tilde c}-({\tilde z}+\tilde c)}(d\rho^2+d\tilde z^2),
\end{eqnarray}
where the coordinate $z$ in the definition of $R_d$ is replaced with $\tilde z$. This coincides with the solution obtained by Elvang and Horowitz~\cite{EH}, which describes static black holes on the Kaluza-Klein bubble.

\subsection{Boosted  black string}
The single boosted black string is achieved by taking the limit $\eta_2 \rightarrow -1$.
This solution corresponds to the electric charged black hole in the effective four
dimensional theory \cite{{Gibbons:1985ac},{Cvetic:1995sz},{Chamblin:1996kw}}.
It can be easily confirmed that the ADM mass (\ref{eq:adm_mass2}) 
and the electric charge (\ref{eq:charge}) become well-known forms
because $\Omega_1=\Omega_2$ when $\eta_2=-1$.
The $3\times3$ metric functions become the following form
\begin{eqnarray}
  g_{tt} &=& -\frac{R_{\lambda\sigma}+R_{\eta_1\sigma}-(2-\lambda-\eta_1)\sigma}
                    {R_{\lambda\sigma}+R_{\eta_1\sigma}+(\lambda-\eta_1)\sigma}, \\
  g_{t\phi} &=&   -\frac{2\sqrt{(1-\lambda)(1-\eta_1)}\sigma}
                    {R_{\lambda\sigma}+R_{\eta_1\sigma}+(\lambda-\eta_1)\sigma}, \\
   g_{\phi\phi} &=&  \frac{R_{\lambda\sigma}+R_{\eta_1\sigma}+(2-\lambda-\eta_1)\sigma}
                    {R_{\lambda\sigma}+R_{\eta_1\sigma}+(\lambda-\eta_1)\sigma}, \\
 g_{\psi\psi} &=& \rho^2 \frac{R_{\lambda\sigma}+R_{\eta_1\sigma}+(\lambda-\eta_1)\sigma}
                             {R_{\lambda\sigma}+R_{\eta_1\sigma}-(\lambda-\eta_1)\sigma}.
\end{eqnarray}
Introducing the Schwarzschild radial coordinate as
\begin{equation}
 r_s = \frac{R_{\lambda\sigma}+R_{\eta_1\sigma}}{2}+m,
\end{equation}
we can derive a familiar expression of the boosted black string from the solitonic solution.

\section{Summary and Discussion}\label{sec:final}

Using the solitonic solution generating methods, we generated a new exact solution which describes a pair of boosted black holes in the compact direction on a Kaluza-Klein bubble as a vacuum solution in the five-dimensional Kaluza-Klein theory. 
This solution cannot be obtained by the simple boost transformation of the static
black holes on Kaluza-Klein bubble.
We also investigated the properties of this solution, particularly, its asymptotic structure, the geometry of the black hole horizons and the Kaluza-Klein bubble and the limits of single black string and static case.
The asymptotic structure is the $S^1$ bundle over the four-dimensional Minkowski spacetime. Two black holes have the topological structure of $S^3$ and the bubble is topologically $S^1\times R$. The solution describes the physical situation such that two black holes have the boost velocity of the same direction and the bubble plays a role in holding two black holes. The ADM mass and the linear momentum of the solution can be written by the two boosted velocity parameters.
In the static case, it coincides with the solution found by Elvang and Horowitz. The solution has a limit of single boosted black string.

In this article, we concentrated on the black hole solution with a linear momentum component. The solution with an angular momentum component has been derived in the previous paper \cite{Tomizawa:2007mz}.
The investigation on the solution with these two components is enormously challenging.
In general, the inverse scattering method can generate a solution with two momentum components. 
We will give such a solution in our future article.

\section*{Acknowledgments}
We thank Ken-ichi~Nakao for continuous encouragement.
This work is partially supported by Grant-in-Aid for Young Scientists (B)
(No. 17740152) from Japanese Ministry of Education, Science,
Sports, and Culture.

\appendix
\section{Solutions generated by ISM}\label{app:ISM}
Following the techniques in the Ref~\cite{Tomizawa, Tomizawa2, Tomizawa3}, we construct a new Kaluza-Klein black hole solution. We consider the five-dimensional stationary and axisymmetric vacuum spacetimes which admit three commuting Killing vectors $\partial/\partial t$, $\partial/\partial \phi$ and $\partial/\partial \psi$, where $\partial/\partial t$ is a Killing vector field associated with time translation, $\partial/\partial \phi$ and $\partial/\partial \psi$ denote spacelike Killing vector fields with closed orbits. In such a spacetime, the metric can be written in the canonical form as
\begin{eqnarray}
ds^2=g_{ij}dx^idx^j+f(d\rho^2+dz^2),
\end{eqnarray}
where the metric components $g_{ij}$ and the metric coefficient $f$ are functions which depend on $\rho$ and $z$ only. The metric $g_{ij}$ satisfies the supplementary condition ${\rm det}\ g_{ij}=-\rho^2$.
We begin with the following seed
\begin{eqnarray}
ds^2=-\frac{R_{\eta_2\sigma}+z-\eta_2\sigma}{R_{\eta_1\sigma}+z-\eta_1\sigma}dt^2+\frac{R_{\lambda\sigma}+z-\lambda\sigma}{R_{\eta_2\sigma}+z-\eta_2\sigma}d\phi^2+\frac{R_{\eta_1\sigma}+z-\eta_1\sigma}{R_{\lambda\sigma}+z-\lambda\sigma}\rho^2d\psi^2+f(d\rho^2+dz^2), \nonumber \\ \label{eq:seed1} 
\end{eqnarray}
where $R_d$ is defined as $R_d:=\sqrt{\rho^2+(z-d)^2}$. The parameters $\eta_1,\eta_2$ and $\lambda$ satisfy the inequality $\eta_1<\eta_2<\lambda<1$ and $\sigma>0$. 
Instead of solving the L-A pair for the seed metric (\ref{eq:seed1}), it is sufficient to consider the following metric form
\begin{eqnarray}
ds^2=-dt^2+g_2d\phi^2+g_3d\psi^2+f(d\rho^2+dz^2),\label{eq:g23}
\end{eqnarray}
where $g_2$ and $g_3$ are given by
\begin{eqnarray}
g_2=\frac{(R_{\eta_1\sigma}+z-\eta_1\sigma)(R_{\lambda\sigma}+z-\lambda\sigma)}{(R_{\eta_2\sigma}+z-\eta_2\sigma)^2},\quad g_3=\frac{(R_{\eta_2}+z-\eta_2\sigma)^2\rho^2}{(R_{\eta_1\sigma}+z-\eta_1\sigma)(R_{\lambda\sigma}+z-\lambda\sigma)}.\label{eq:g23-2}
\end{eqnarray}
Let us consider the conformal transformation of the two dimensional metric $g_{AB}\ (A,B=t,\phi)$ and the rescaling of the $\psi\psi$-component in which the determinant ${\rm det} g$ is invariant
\begin{eqnarray}
g_{0}={\rm diag} (-1,g_2,g_3)\to g'_0={\rm diag} (-\Omega,\Omega g_2,\Omega^{-2}g_3),\label{eq:conf}
\end{eqnarray} 
where $\Omega$ is the $tt$-component of the seed (\ref{eq:seed1}), i.e. 
\begin{eqnarray}
\Omega=\frac{R_{\eta_2\sigma}+z-\eta_2\sigma}{R_{\eta_1\sigma}+z-\eta_1\sigma}.
\end{eqnarray} 
Then, under this transformation, the three-dimensional metric coincides with the metric (\ref{eq:seed1}). On the other hand, as discussed in~\cite{Tomizawa3}, under this transformation the physical metric of two-solitonic solution is transformed as
\begin{eqnarray}
g=\left(
\begin{array}{@{\,}c|ccc@{\,}}
\displaystyle
g_{AB}
& 0 \\ \hline 0  &
g_3
\end{array}
\right)
\to 
g^{\prime}=\left(
\begin{array}{@{\,}c|ccc@{\,}}
\displaystyle
\Omega g_{AB}
& 0 \\ \hline 0  &
\Omega^{-2}g_3
\end{array}
\right).\label{eq:tr}
\end{eqnarray}
This is why  we may perform the transformation (\ref{eq:conf}) for the two-solitonic solution generated from the seed (\ref{eq:g23}) in order to obtain the two-solitonic solution from the seed (\ref{eq:seed1}). The generating matrix $\psi_0$ for this seed metric (\ref{eq:g23}) is computed as follows
\begin{eqnarray}
\psi_0[\bar \lambda]={\rm diag}\left(-1,\psi_2[\bar \lambda],\psi_3[\bar \lambda]\right)\nonumber
\end{eqnarray}
with
\begin{eqnarray*}
& & \psi_2[\bar \lambda]=\frac{(R_{\eta_1\sigma}+z-\eta_1\sigma+\bar \lambda)(R_{\lambda\sigma}+z-\lambda\sigma+\bar\lambda)}{(R_{\eta_2\sigma}+z-\eta_2\sigma+\bar\lambda)^2},\nonumber\\
& &\psi_3[\bar\lambda]=\frac{(R_{\eta_2\sigma}+z-\eta_2\sigma+\bar\lambda)^2(\rho^2-2z\bar\lambda-{\bar\lambda}^2)}{(R_{\eta_1\sigma}+z-\eta_1\sigma+\bar\lambda)(R_{\lambda\sigma}+z-\lambda\sigma+\bar\lambda)}.
\end{eqnarray*}
Then, the two-solitonic solution is obtained as
\begin{eqnarray*}
& &g^{{\rm (phys)}}_{tt}=
-\frac{\Omega G_{tt}}{\mu_1\mu_2\Sigma},\quad
g_{t\phi}^{{\rm (phys)}}
=-g_2\frac{\Omega(\rho^2+\mu_1\mu_2)
G_{t\phi}}{\mu_1\mu_2 \Sigma},\quad
g^{{\rm (phys)}}_{\phi\phi}=
-g_2\frac{\Omega G_{\phi\phi}}{\mu_1\mu_2\Sigma},
\label{eq:gphys}
\\
& &g^{{\rm (phys)}}_{\psi\psi}=\Omega^{-2}g_3,\quad
g_{\phi\psi}^{{\rm (phys)}}=g_{t\psi}^{{\rm (phys)}}=0,
\end{eqnarray*}
where the functions $G_{tt},\ G_{t\phi},\ G_{\phi\phi}$ and $\Sigma$ are given by
\begin{eqnarray}
G_{tt}&=&-m_{01}^{(1)2}m_{01}^{(2)2}
\psi_2[\mu_1]^2\psi_2[\mu_2]^2(\mu_1-\mu_2)^2
\rho^4+m_{01}^{(1)2}m_{02}^{(2)2}g_2\mu_2^2
(\rho^2+\mu_1\mu_2)^2\psi_2[\mu_1]^2
\nonumber\\
& &+m_{01}^{(2)2}m_{02}^{(1)2}g_2 \mu_1^2
(\rho^2+\mu_1\mu_2)^2\psi_2[\mu_2]^2
-m_{02}^{(1)2}m_{02}^{(2)2}g_2^2
\mu_1^2\mu_2^2(\mu_1-\mu_2)^2\\
& &-2m_{01}^{(1)}m_{01}^{(2)}m_{02}^{(1)}m_{02}^{(2)}
g_2\psi_2[\mu_1]\psi_2[\mu_2](\rho^2+\mu_1^2)
(\rho^2+\mu_2^2)\mu_1\mu_2,
\nonumber
\end{eqnarray}
\begin{eqnarray}
G_{\phi\phi}&=&m_{01}^{(1)2}m_{01}^{(2)2}
\mu_1^2\mu_2^2(\mu_1-\mu_2)^2
\psi_2[\mu_1]^2\psi_2[\mu_2]^2
+m_{02}^{(1)2}m_{02}^{(2)2}g_2^2
(\mu_1-\mu_2)^2\rho^4\nonumber\\
& &-m_{01}^{(1)2}m_{02}^{(2)2}g_2\mu_1^2
\psi_2[\mu_1]^2(\rho^2+\mu_1\mu_2)^2
-m_{01}^{(2)2}m_{02}^{(1)2}g_2\mu_2^2
\psi_2[\mu_2]^2(\rho^2+\mu_1\mu_2)^2  \\
& &+2m_{01}^{(1)}m_{01}^{(2)}m_{02}^{(1)}m_{02}^{(2)}
g_2\mu_1\mu_2\psi_2[\mu_2]\psi_2[\mu_1]
(\rho^2+\mu_1^2)(\rho^2+\mu_2^2),
\nonumber
\end{eqnarray}
\begin{eqnarray}
G_{t\phi}&=&m_{01}^{(1)}m_{01}^{(2)2}
m_{02}^{(1)}\mu_2(\mu_1-\mu_2)
\psi_2[\mu_2]^2\psi_2[\mu_1](\rho^2+\mu_1^2)\nonumber\\
& &+m_{01}^{(1)}m_{02}^{(1)}m_{02}^{(2)2}
g_2\mu_2(\mu_2-\mu_1)
\psi_2[\mu_1](\rho^2+\mu_1^2)\nonumber\\
& &+m_{01}^{(1)2}m_{01}^{(2)}m_{02}^{(2)}
\mu_1(\mu_2-\mu_1)\psi_2[\mu_1]^2
\psi_2[\mu_2](\rho^2+\mu_2^2)\nonumber\\
&&+m_{01}^{(2)}m_{02}^{(1)2}m_{02}^{(2)}
\mu_1g_2\psi_2[\mu_2](\rho^2+\mu_2^2)(\mu_1-\mu_2),
\end{eqnarray}

\begin{eqnarray}
\Sigma&=&m_{01}^{(1)2}m_{01}^{(2)2}
\psi_2[\mu_1]^2\psi_2[\mu_2]^2(\mu_1-\mu_2)^2\rho^2
+m_{02}^{(1)2}m_{02}^{(2)2}g_2^2(\mu_1-\mu_2)^2\rho^2\nonumber\\
& &+m_{01}^{(1)2}m_{02}^{(2)2}g_2\psi_2[\mu_1]^2
(\rho^2+\mu_1\mu_2)^2
+m_{02}^{(1)2}m_{01}^{(2)2}g_2
\psi_2[\mu_2]^2(\rho^2+\mu_1\mu_2)^2\nonumber\\
& &-2m_{01}^{(1)}m_{01}^{(2)}m_{02}^{(1)}m_{02}^{(2)}
g_2\psi_2[\mu_1]\psi_2[\mu_2](\rho^2+\mu_1^2)(\rho^2+\mu_2^2).
\end{eqnarray}
Here, $\mu_1$ and $\mu_2$ are given by
\begin{eqnarray}
\mu_1(\rho,z)=\sqrt{\rho^2+(z+\sigma)^2}-(z+\sigma),\quad \mu_2(\rho,z)=\sqrt{\rho^2+(z-\sigma)^2}-(z-\sigma).
\end{eqnarray}
We should note that this three-dimensional metric $g^{{\rm (phy)}}_{ij}$ satisfies the supplementary condition ${\rm det}\ g_{ij}=-\rho^2$.  
Next, let us consider the coordinate transformation
of the physical metric such that
\begin{eqnarray}
t\rightarrow t'=t-C_1\phi, \qquad \phi
\rightarrow \phi'=\phi,
\end{eqnarray}
where $C_1$ is a constant. Under this transformation, the physical metric  becomes
\begin{eqnarray}
& &g_{tt}^{\rm (phys)}\rightarrow
g_{tt}=g_{tt}^{\rm (phys)},
\nonumber \\
& &g_{t\phi}^{\rm (phys)}\rightarrow
g_{t\phi}=g_{t\phi}^{\rm (phys)}+C_1
g_{tt}^{\rm (phys)},\label{eq:solution}\\ 
& &g_{\phi\phi}^{\rm (phys)}\rightarrow
g_{\phi\phi}=g_{\phi\phi}^{\rm (phys)}
+2C_1 g_{t\phi}^{\rm (phys)}+C_1^2g_{tt}^{\rm (phys)}.
\nonumber
\end{eqnarray}
Here, we should note that the transformed metric also satisfies the
supplementary condition ${\rm det} g=-\rho^2$. Though the metric seems to contain the four new parameters $m_{01}^{(1)},m_{01}^{(2)},m_{02}^{(1)}$ and $m_{02}^{(2)},$ it can be written only in term of the ratios
\begin{eqnarray}
\alpha:=\frac{m_{02}^{(2)}}{m_{01}^{(2)}},\quad \beta:=-\frac{m_{01}^{(1)}}{m_{02}^{(1)}}. 
\end{eqnarray}
Using  the parameters $\alpha$ and $\beta$, we can write all components of the metric. The metric function  $f(\rho,z)$ takes the following form
\begin{eqnarray}
f=\frac{C_2Y_{\sigma,-\sigma}Y_{-\sigma,\eta_2\sigma}}{16\sigma^2Y_{\sigma,\eta_2\sigma}}\sqrt{\frac{Y_{\sigma,\eta_1\sigma}Y_{\sigma,\lambda\sigma}Y_{\eta_1\sigma,\eta_2\sigma}Y_{\lambda\sigma,\eta_1\sigma}Y_{\lambda\sigma,\eta_2\sigma}}{Y_{-\sigma,-\sigma}Y_{\eta_1\sigma,\eta_1\sigma}Y_{\eta_2\sigma,\eta_2\sigma}Y_{\lambda\sigma,\lambda\sigma}Y_{-\sigma,\eta_1\sigma}Y_{-\sigma,\lambda\sigma}Y_{\sigma,\sigma}}}\frac{\Omega Y}{(\rho^2+\mu_1\mu_2)^4\mu_1^3\mu_2\psi_2[\mu_2]^2}, \nonumber \\ \label{eq:f}
\end{eqnarray}
where $C_2$ is an arbitrary constant, $Y_{c,d}$ is defied as $Y_{c,d}:=R_cR_d+(z-c)(z-d)+\rho^2$ and the function $Y$ is given by
\begin{eqnarray}
Y&=&\rho^2[16\beta\sigma^2\mu_1^2\mu_2^2\psi_2[\mu_1]\psi_2[\mu_2]-\alpha g_2(\mu_1-\mu_2)^2(\rho^2+\mu_1\mu_2)^2]^2\nonumber\\
 & &+16\sigma^2g_2\mu_1^2\mu_2^2(\rho^2+\mu_1\mu_2)^4(\psi_2[\mu_2]+\alpha\beta\psi_2[\mu_1])^2.\nonumber
\end{eqnarray}



\begin{thebibliography}{99}

\bibitem{KK}
T. Kaluza, Sitzungsber. Preuss. Akad. Wiss. Berlin (Math. Phys. K) 1, 966 (1921);
O. Klein, Z. Phys. 37, 895 (1926).

\bibitem{Overduin:1998pn}
  J.~M.~Overduin and P.~S.~Wesson,
  Phys.\ Rept.\  {\bf 283}, 303 (1997).


\bibitem{IM}
H. Ishihara and K. Matsuno, Prog.Theor.Phys. {\bf 116}, 417 (2006).

\bibitem{Ishihara}
H. Ishihara, M. Kimura, K. Matsuno and  S. Tomizawa, Class. Quant. Grav. {\bf 23}, 6919 (2006).


\bibitem{Elvang:2004iz}
  H.~Elvang, T.~Harmark and N.~A.~Obers,
  JHEP {\bf 0501}, 003 (2005).






\bibitem{Witten}
E. Witten, Nucl. Phys. {\bf B195}, 481 (1982). 

\bibitem{weyl}
R. Emparan, H. S. Reall, Phys. Rev. D {\bf 65}, 084025 (2002). 

\bibitem{EH}
H. Elvang and G. T. Horowitz, Phys. Rev. D {\bf 67}, 044015 (2003).

\bibitem{Tomizawa:2007mz}
  S.~Tomizawa, H.~Iguchi and T.~Mishima,
  arXiv:hep-th/0702207.


\bibitem{Harrison}
B.~K.~Harrison, Phys. Rev. Lett. {\bf 41}, 1197 (1978);\\
Erratum-ibid. Phys. Rev. Lett. {\bf 41}, 1835 (1978).

\bibitem{Neugebauer}
G.~Neugebauer, J.\ Phys.\ A {\bf 13}, L19 (1980).


\bibitem{Belinskii}
V.~A.~Belinskii and V.~E.~Zakharov, Sov. Phys. JETP {\bf 50}, 1 (1979);\\
V.~A.~Belinskii and V.~E.~Zakharov, Sov. Phys. JETP {\bf 48}, 985 (1978);\\
V.~A.~Belinski and E.~Verdaguer, {\it Gravitational Solitons} (CambridgeUniversity Press, Cambridge, England, 2001);\\
H.~Stephani, D.~Kramer, M.~MacCallum, C.~Hoenselaers and E.~Herlt, {\it Exact solutions of Einstein's Field Equations, 2nd ed.} (Cambridge University Press, Cambridge, 2003).

%

\bibitem{Mishima}
T.~Mishima and H.~Iguchi,
Phys. Rev. D {\bf 73}, 044030 (2006);\\
H.~Iguchi and T.~Mishima, Phys. Rev. D {\bf 74}, 024029 (2006). 


\bibitem{MI2}
H. Iguchi and T. Mishima, Phys. Rev. D {\bf 73}, 121501(R) (2006).


\bibitem{MI3}
H. Iguchi and T. Mishima, 
  Phys.\ Rev.\  D {\bf 75}, 064018 (2007).



\bibitem{Koikawa}
T. Koikawa, Prog. Theor. Phys. {\bf 114}, 793 (2005).

\bibitem{Tomizawa}
S. Tomizawa, Y. Morisawa, Y Yasui, Phys. Rev. D{\bf 73}, 064009 (2006).

\bibitem{Azuma}
T. Azuma and T. Koikawa, Prog. Theor. Phys. {\bf 116}, 319 (2006).

\bibitem{Pomeransky:2005sj}
A.~A.~Pomeransky, Phys. Rev. D {\bf 73}, 044004 (2006). 

\bibitem{Tomizawa2}
S. Tomizawa and M. Nozawa, Phys. Rev. D {\bf 73}, 124034 (2006). 

\bibitem{Pomeransky2}
A. A. Pomeransky, R. A. Sen'kov, arXiv:hep-th/0612005. 

\bibitem{EF}
H. Elvang and P. Figueras, arXiv:hep-th/0701035.

\bibitem{Tomizawa3}
S. Tomizawa, H. Iguchi and T. Mishima,  Phys. Rev. D {\bf 74},104004 (2006).

\bibitem{Dobiasch:1981vh}
  P.~Dobiasch and D.~Maison,
  Gen.\ Rel.\ Grav.\  {\bf 14}, 231 (1982).


\bibitem{Gibbons:1985ac}
  G.~W.~Gibbons and D.~L.~Wiltshire,
  Annals Phys.\  {\bf 167}, 201 (1986).

\bibitem{Cvetic:1995sz}
  M.~Cvetic and D.~Youm,
  Phys.\ Rev.\ Lett.\  {\bf 75}, 4165 (1995).



\bibitem{Chamblin:1996kw}
  A.~Chamblin and R.~Emparan,
  Phys.\ Rev.\  D {\bf 55}, 754 (1997).


\bibitem{Kastor:2007wr}
  D.~Kastor, S.~Ray and J.~Traschen,
  arXiv:0704.0729 [hep-th].



\bibitem{Castejon-Amenedo:1990b}
J.~Castejon-Amenedo and V.~S.~Manko, Phys.\ Rev.\ D {\bf 41}, 2018 (1990).


%
%
%
%


\bibitem{Harmark}

T. Harmark, Phys. Rev. D {\bf 70}, 124002 (2004).


























 







\end{thebibliography}
\end{document}